\documentclass[aps,prl,preprint,superscriptaddress]{revtex4}
\usepackage{graphicx}
\usepackage{color}
\usepackage{amsfonts}

\begin{document}
\title{Impact of Edge States on Device Performance of Phosphorene Heterojunction Tunneling Field Effect Transistors}
\author{Fei Liu}
\email{fliu003@gmail.com}
\affiliation{Department of Physics, The University of Hong Kong, Hong Kong, China}
\affiliation{Department of Physics, McGill University, Montreal, H3A 2T8, Canada}

\author{Jian Wang}
\affiliation{Department of Physics, The University of Hong Kong, Hong Kong, China}

\author{Hong Guo}
\affiliation{Department of Physics, McGill University, Montreal, H3A 2T8, Canada}

\begin{abstract}
Black phosphorus (BP) tunneling transistors (TFETs) using heterojunction (He) are investigated by atomistic quantum transport simulations. It is observed that edge states have a great impact on  transport characteristics of BP He-TFETs, which result in the potential pinning effect and deteriorate the gate control. While, on-state current can be effectively enhanced by using  hydrogen to saturate the edge dangling bonds in BP He-TFETs, in which edge states are quenched. By extending layered BP with a smaller band gap to the channel region and modulating the BP thickness, device performance of BP He-TFETs can be further optimized and fulfill the requirements of the international technology road-map for semiconductors (ITRS) 2013 for low power applications. In 15 nm 3L-1L and 4L-1L BP He-TFETs along armchair direction on-state current can reach above 10$^3$ $\mu$A/$\mu$m with the fixed off-state current of 10 $pA/\mu$m. It is also found that ambipolar effect can be effectively suppressed in BP He-TFETs. 
\end{abstract}
\newpage
\maketitle

With continuous device scaling it becomes more and more important to reduce power consumption in integrated circuits, which can be realized by reducing the supply voltage. For this purpose, transistors should achieve a lower subthreshold swing (SS).  In traditional CMOS technology, metal-oxide-semiconductor field effect transistors (MOSFETs) are controlled by manipulating the thermionic current over the barrier and SS can not be smaller than 60 mV/decade at room temperature\cite{ACSeabaugh,AMIonescu}.
In recent years, tunneling field effect transistors(TFETs) are extensively studied due to the potential to overcome the thermal sub-threshold limit\cite{ACSeabaugh,AMIonescu}. With smaller SS and lower off-state current both static and dynamic power consumptions can be effectively decreased in TFETs. More recently, two-dimensional (2D) structures including graphene \cite{PZhao,FWChen}, transition metal dichalcogenides (TMDCs)\cite{YYoon,NMa,LTLam,FLiu} and topological insulators\cite{JChang,QZhang}, are explored as the channel materials of TFETs for good gate control. However, these TFETs also have their limitations. Inevitable edge roughness increases the off-state current of GNRs TFETs\cite{MLuisier}; it is hard to get reasonable on-state current in TMDCs \cite{LTLam,FLiu}and TI TFETs\cite{JChang,QZhang} for realistic low power applications.

At the beginning of 2014, layered black phosphorus (BP) with  many unique properties is discovered and applied in nanoelectronic and nanophotonic devices\cite{LLi,HLiu,WangH,ZhuW,HaratipourN,XiaF,EngelM}. Different from other 2D semiconductors, BP has a layer dependent direct band gap and anisotropic band structure\cite{JQiao,VTran,FeiR,SBZhang}. The unique property of anisotropy can provide a small effective mass and large  density of states at the same time. Therefore, phosphorene TFETs can reach higher on-state current than TMDCs TFETs\cite{JWChang,FLiu02}. On-current can be boosted by using multilayer BP films, while leakage current and subthreshold swing(SS) is increased at the same time\cite{FLiu02}. Due to the layer dependent band gap, it should be experimentally feasible to design heterojunction by thickness modulation for performance optimization in BP TFETs.

In this work, we studied  the device physics of BP heterojunction (He) TFETs through atomistic quantum transport simulations, and compared device performance of BP He-TFETs with BP homojunction (Ho) TFETs. Thicker BP with a smaller band gap is utilized in the source to obtain higher on-state current and thinner BP is applied in the channel and drain to achieve lower off-state current. We perform the first principles calculations to study the edge states, and demonstrate that these interface sates have a great impact on transport properties of BP He-TFETs. Motivated by the anisotropic band structure He-TFETs along armchair direction (AD) and zigzag direction (ZD) are examined. For optimizing device performance, thickness of BP film is modulated and layered BP applied in the source is extended to the channel. Finally, scale behavior of BP He-TFETs is studied according to low power technology requirements specified in ITRS 2013\cite{ITRS}.

\begin{figure*}[t!]
\centering
\includegraphics[width=5.5in]{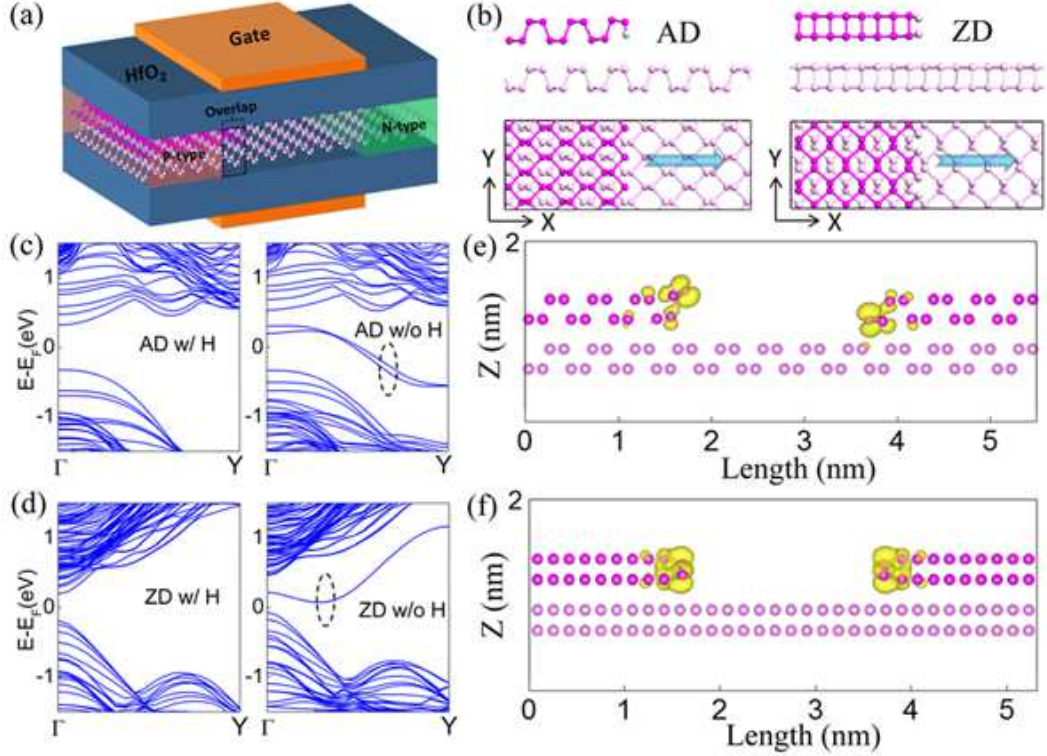}
\caption{(a) Schematic illustration of a BP double gate TFET using heterojunction with 10 nm channel length.  Bilayer BP is applied in the source, the channel and the drain is monolayer BP. There is a overlap region of bilayer BP in the dashed box under the gate which is also intrinsic. (b) Atomistic structures of BP heterojunctions with hydrogen passivated edges:  armchair direction (AD, left panel) and zigzag direction (ZD, right panel). Band structures of periodic BP heterojunctions with/without hydrogen passivation: (c) AD and (d) ZD. Local charge densities of edge bands (marked by dash circle lines in (c) and (d)) of periodic BP heterojunctions without hydrogen passivation: (e) AD and (f) ZD. }\label{Fig01}
\end{figure*}

As illustrated in Fig. \ref{Fig01}(a), the simulated device has a double gate structure with BP heterojunction sandwiched between two 3 nm HfO$_2$ layers. The source is  heavily doped to p-type with the doping density of n$_0$ = 7.0 $\times 10^{13}/cm^{-2}$, and the drain is doped to n-type with the same density. The intrinsic channel length of the TFET is equal to the length of the gate. In homojunction, source, drain and the channel under the gate use  the same layer BP. While, in heterojunction bilayer BP is applied in the source and monolayer BP is applied in the channel and drain as illustrated in Fig. \ref{Fig01}(a). In order to improve device performance a overlap region (dashed box region in Fig. \ref{Fig01}(a)) with length L$_{OL}$ near the source under the gate has been considered, which has the same number of layers with the source.

We first study the electronic properties of BP heterojunctions. The first principles calculations are performed within the density functional theory (DFT) implemented in the Vienna ab initio simulation package (VASP)\cite{GKresse}. The generalized gradient approximation (GGA) with the Perdew-Burke-Ernzerhof (PBE) functional is applied to treat the exchange-correlation\cite{PBE01,PBE02}. Fig. \ref{Fig01}(b) shows the optimized atomic structures of 2L-1L heterojunctions using VASP with hydrogen (H) passivated edges. The DFT-calculated band structures of 2L-1L periodic structures with/without H passivation as demonstrated in Fig. \ref{Fig01}(e, f)  are presented in Fig. \ref{Fig01}(c, d). These atomistic structures are also optimized by using VASP. It is obviously found that there are interface states in BP heterojunction without H passivation in both AD and ZD due to the edge dangling bonds. These interface states localized at the heterojunction edges as illustrated in Fig. \ref{Fig01}(e, f) and can be quenched by H atoms shown in left panels of Fig. \ref{Fig01}(c, d).  While, BP heterojunctions in the two directions have different electronic properties. Due to the edge dangling bonds AD BP heterojunction without H passivation is metallic with the Fermi energy across the edge bands; while ZD BP heterojunction is semiconducting.

\begin{figure*}[t!]
\centering
\includegraphics[width=6.4in]{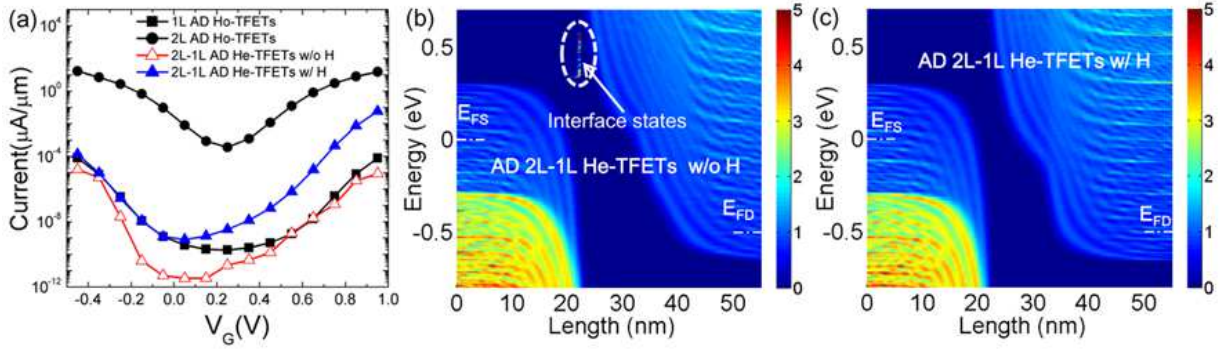}
\caption{(a) $I_D$-$V_G$ of BP Ho-TFETs and He-TFETs in AD with 10 nm gate length at V$_D$ = 0.5 V. Local density states (LDOS) of (b)He-TFETs without hydrogen passivation and (c) He-TFETs with hydrogen passivation at V$_G$ = 0.85 V. }\label{Fig02}
\end{figure*}

Fig. \ref{Fig02}(a) shows the $I_D$ vs $V_G$ characteristics of four kinds of TFETs in AD: 1L BP Ho-TFETs, 2L BP Ho-TFETs and 2L-1L BP He-TFETs with/without edge passivation. Layered BPs are described by four-band tight-binding (TB) model for transport calculations\cite{ANRudenko,FLiu02}. The nearest interlayer coupling parameter is adjusted to fit the GW band structure: t$_1^{\perp}$ = 0.355 eV, 0.398 eV and 0.427 eV for 2L, 3L and 4L. The band gaps of 1L, 2L, 3L and 4L BP are 1.52 eV, 1.01 eV, 0.68 eV and 0.46 eV, respectively. To study the edge passivation with H atoms in AD He-TFETs using the TB model, phosphorus atoms at the interface edges are removed because the passivated phosphorus atoms are not available for carriers\cite{SBhandary}. Ballistic transport of BP TFETs are calculated by solving open-boundary  Schr\"{o}dinger equation and Poisson equation self-consistently within the non-equilibrium Green's function (NEGF) formalism\cite{SDatta}. It is well known that the band gap decreases with the number of BP layers. Band gaps of monolayer and bilayer BP are 1.52 eV and 1.01 eV, respectively\cite{FLiu02}. Therefore, current is greatly enlarged in 2L Ho-TFETs. The off-current of 2L Ho-TFETs in AD at $V_G = V_D/2$ is larger than that of 1L Ho-TFETs in AD by over six orders. Due to the larger band gap 1L Ho-TFETs have greater I$_{on}$/I$_{off}$ ratio of 2.1$\times$10$^3$ and SS is 151 mV/decade, where on-state and off-state are set at V$_{G,off}$ = V$_{D}$/2 and V$_{G,on}$ = V$_{G,off}$+V$_{D}$. Generally, on-state current can be boosted by using smaller band gap material in the source\cite{JKnoch}. Drain current is increased in 2L-1L AD He-TFETs with edge passivation at $V_G$ = 0.85 V; however, the current is reduced in 2L-1L AD He-TFETs without edge passivation as shown in Fig. \ref{Fig02}(a).

\begin{figure}[t!]
\centering
\includegraphics[width=3.0in]{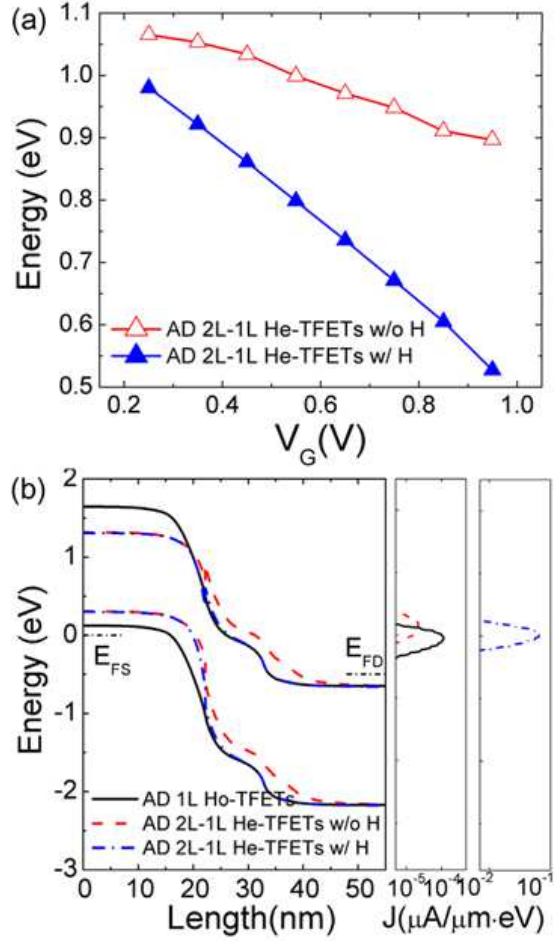}
\caption{(a) Conduction band minimum (CBM) at the source-channel interface as a function of gate voltage. (b) Band edge profiles and energy resolved current spectrums of He-TFETs and Ho-TFETs at V$_G$ = 0.85 V. } \label{Fig03_new}
\end{figure}

To analyze the reason why the on-state current is not effectively improved in  2L-1L AD BP He-TFETs without edge passivation, we calculated local density of states (LDOS) and band profiles along the channel as illustrated in Fig. \ref{Fig02}(b, c). Edge states can be obviously found in 2L-1L BP heterojunction without edge passivation shown in Fig. \ref{Fig02}(b); while, these edge states can be eliminated in H saturated BP He-TFETs in Fig. \ref{Fig02}(c), which is consistent with DFT results. Due to the existence of  edge states the channel potential is pinned in 2L-1L AD BP He-TFETs without edge passivation. Therefore, gate voltage can not effectively modulate the channel potential. Fig. \ref{Fig03_new}(a) demonstrates the conduction band minimum (CBM) at source-channel junction where  edge states appear in the two kinds of devices. It can be found that gate control is deteriorated in BP He-TFETs with edge states which lead to potential pinning effect. When gate voltage is increased from 0.25 V to 0.95 V, the CBM is pushed down by 0.45 eV and 0.17 eV  for BP He-TFETs with and without edge passivation, respectively.

For the existence of the edge states the electronic properties in the nearby of 2L-1L interface are actually different from left semi-infinite lead. From Fig. \ref{Fig03_new}(b), 2L-1L AD BP He-TFETs without H passivation have a longer tunneling length of 10.8 nm at E = 0 eV than 10 nm in 1L AD BP Ho-TFETs.  Even though 2L BP has a smaller hole effective mass in AD than 1L BP: m$_{AD,2L}$ = 0.16 m$_0$ and m$_{AD,1L}$ = 0.18 m$_0$, the current density at $V_G = 0.85 V$  gets smaller as shown in Fig. \ref{Fig03_new}(b) and $I_{on}$ is not improved in 2L-1L AD BP He-TFETs without H passivation compared with 1L BP Ho-TFETs. Due to the longer tunneling length at the source and drain Fermi levels, 2L-1L AD He-TFETs without edge passivation have smaller current than 1L AD Ho-TFETs at negative gate voltages.

In contrast, the drain current  is effectively increased from $8.6 \times 10^{-6}$ $\mu A/\mu m$ in 1L AD BP Ho-TFETs to $7.4 \times 10^{-3}$ $\mu A/\mu m$ at $V_G = 0.85 V$. There is no edge state in 2L-1L AD BP He-TFETs with H passivation as shown in Fig. 2(c) and the tunneling length   at $V_G = 0.85 V$ gets thinner than 1L AD Ho-TFETs in Fig. \ref{Fig03_new}(b).  At $V_G = 0.85 V$, the current mainly depends on the tunneling from source valence band (VB) to channel conduction band (CB) and is determined by properties of source-channel interface. 2L BP applied in the source of BP He-TFETs leads to more prominent band bending as shown in Fig. \ref{Fig03_new}(b). Both the prominent band bending and the smaller band gap in BP He-TFETs result in smaller tunneling height and narrower tunneling width from source VB to channel CB. Hence, current is greatly increased as demonstrated in Fig. \ref{Fig03_new}(b). At $V_G = 0.25 V$, the current is contributed by direct tunneling from source VB to drain CB. So, the tunneling process is mainly dominated by the properties of the channel material: tunneling barrier height related the band gap and tunneling length. In 2L-1L AD BP He-TFETs the channel material is 1L BP with a band gap of 1.52 eV, so the off-state current can be kept very small even though larger than 1L BP Ho-TFETs. Another interesting phenomenon in 2L-1L BP He-TFETs is that ambipolar effect is greatly suppressed. Namely, current at negative gate voltages is extremely smaller than that at high positive gate voltages as shown in Fig. \ref{Fig02}(a). The reason is that we applied an asymmetry device structure. At negative gate voltages the current is determined by the channel-drain junction which is monolayer BP.

\begin{figure}[t!]
\centering
\includegraphics[width=5in]{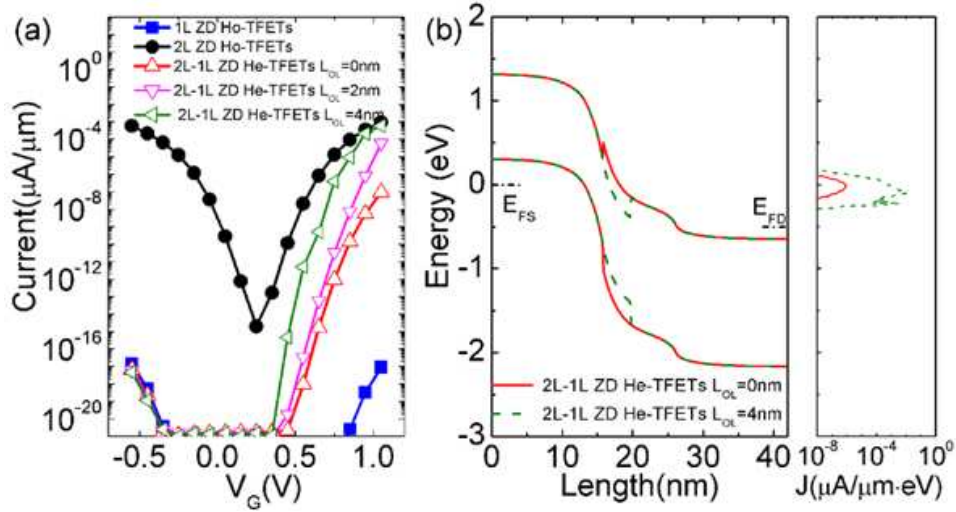}
\caption{(a)  $I_D$-$V_G$  of 10 nm BP Ho-TFETs and He-TFETs in ZD with edge passivation and different overlap lengths (L$_{OL}$) at V$_D$ = 0.5 V. (b) Band edge profiles and energy resolved current spectrums of He-TFETs and Ho-TFETs in ZD at V$_G$ = 1.05 V. } \label{Fig03}
\end{figure}

It is well known that layered BP has anisotropic effective mass; therefore, orientation dependent transport properties in BP He-TFETs are expected. Fig. \ref{Fig03}(a) compares  $I_D$ - $V_G$ characteristics of 1L, 2L ZD BP Ho-TFETs and 2L-1L ZD BP He-TFETs without the overlap region. BP TFETs in ZD show more prominent layer dependent transfer characteristics. As the BP thickness changes from monolayer to bilayer, $I_D$ is increased by over ten orders of magnitude at all studied gate voltages. BP TFETs show orientation dependent transport properties, and $I_D$ of Ho-TFETs in ZD is smaller than that in AD by orders of magnitude for heavier carrier effective masses in ZD\cite{FLiu02}. Compared with 1L BP Ho-TFETs in ZD, 2L-1L ZD BP He-TFETs have larger on-state current and maintain low off-state current. Therefore, SS of BP He-TFETs in ZD is effectively decreased and can reach 40 mV/decade.

We note that even though current at high positive gate voltages is improved in ZD He-TFETs, the on-state current is still smaller than that of 2L BP Ho-TFETs by several orders. In order to further optimize device characteristics 2L BP is extended into the channel. As shown in Fig. \ref{Fig03}(a), the overlap region can greatly improve the on-state current. The current at $V_G = 1.05 V$ is increased by over  three orders of magnitude in 2L-1L BP He-TFETs with $L_{OL} = 4nm$ compared with BP He-TFETs without the overlap region.  The reason is that the extended 2L BP under the gate reduces both the tunneling barrier height and tunneling width, and source-channel junction in the He-TFETs at high gate voltages is the same as that in 2L ZD BP Ho-TFETs  as shown in Fig. \ref{Fig03}(b). So, the on-state current can be as large as that in 2L ZD BP Ho-TFETs.

\begin{figure}[t!]
\centering
\includegraphics[width=4.5in]{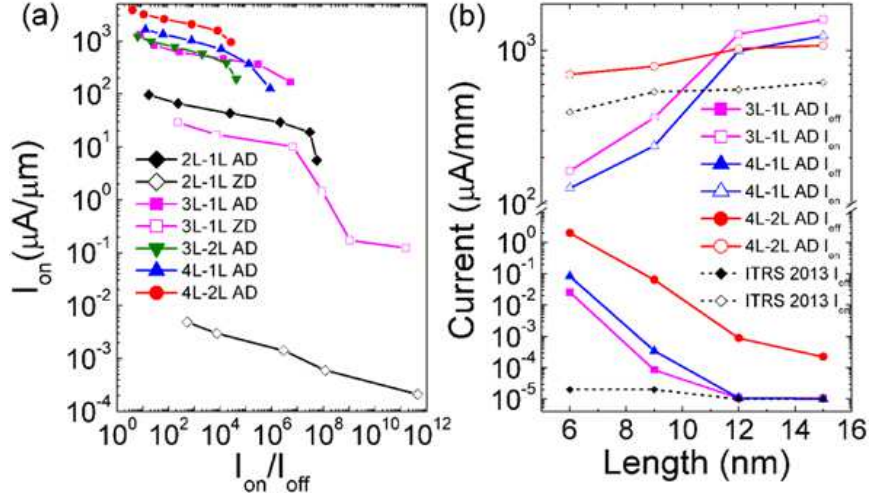}
\caption{(a)  I$_{on}$ as a function of I$_{on}$/I$_{off}$ ratio for 10 nm Ho-TFETs and He-TFETs with edge passivation and $L_{OL}$ = 4 nm at V$_D$ = 0.5 V. (b) I$_{on}$ and I$_{off}$ as a function of channel length of BP AD He-TFETs with $L_{OL}$ = 2 nm. According to ITRS 2013 different drain voltages are applied: V$_{D}$ = 0.68 V, 0.74 V, 0.78 V and 0.83 V for 6 nm, 9 nm , 12 nm and 15 nm BP TFETs, respectively.}\label{Fig04}
\end{figure}

At last, we evaluate the I$_{on}$ as a function of I$_{on}$/I$_{off}$ ratio for BP He-TFETs with H passivated edges and 4 nm overlap region as shown in Fig. \ref{Fig04}(a). It can be found that ZD BP He-TFETs have a larger utmost limit of I$_{on}$/I$_{off}$ ratio than AD BP He-TFETs but smaller on-state current.  In those devices with I$_{on}$/I$_{off}$ ratio larger than 10$^6$  3L-1L  AD He-TFETs have the largest $I_{on}$ of 167$\mu$A/$\mu$m with I$_{on}$/I$_{off}$ ratio of 5.2$\times$ 10$^6$. By increasing the BP film thickness in AD He-TFETs the on-state current can be increased, however utmost limit of I$_{on}$/I$_{off}$ ratio is reduced too. In 4L-2L AD He-TFETs I$_{on}$ can be larger than 10$^3$ $\mu$A/$\mu$m with I$_{on}$/I$_{off}$ ratio larger than 10$^4$. To evaluate BP He-TFETs performance for low power applications according to ITRS 2013\cite{ITRS}, we extracted $I_{on}$ and $I_{off}$ of 3L-1L, 4L-1L and 4L-2L AD BP He-TFETs with different gate lengths as shown in Fig. \ref{Fig04}(b). At the channel lengths of  12 nm and 15nm the minimum current of 4L-1L and 3L-1L AD BP He-TFETs can be less than $10^{-5}$ $\mu A/\mu m$ as required in ITRS 2013 for low power applications; while, when the channel length reaches 9nm and 6nm the minimum current can not meet the off-state current requirement of ITRS 2013. The minimum current of 4L-2L AD BP He-TFETs is always larger than $10^{-5}$ $\mu A/\mu m$ at all simulated channel lengths. Hence, for those devices with the minimum drain current less than that in ITRS 2013 the off current is fixed at the value specified in ITRS 2013; otherwise, the off current is obtained at $V_{G,off} = V_D/2$. On-state current is obtained at $V_{G,on} = V_{G,off}+V_D$. It is found that on-state currents of 3L-1L and 4L-2L AD He-TFETs with H saturated edges at $L_G$ = 12 nm and 15 nm can meet the requirement of ITRS 2013 for low power applications with fixed $I_{off}$ = 10 pA/$\mu$m. The on-state current of 15nm 3L-1L AD He-TFETs reaches $1.6 \times 10^3 \mu A/\mu m$.

In summary, device physics of BP heterojunction tunneling FETs is studied by atomistic simulations. It is discovered that edge states have a great impact on device characteristics of BP He-TFETs, which lead to the potential pinning effect and gate control deterioration. While, on-state current and on-off current ratio can be effectively improved in BP He-TFETs with hydrogen saturated edges. Device performance of BP He-TFETs can be further optimized by extending the low gap material to the channel region and using thicker BP in the source. It is also observed that ambipolar effect can de effectively suppressed in BP He-TFETs due to the asymmetry device structure. Compared with ITRS 2013, 12 nm and 15 nm BP He-TFETs have promising performance for low power applications. \\
 \\

\section{Acknowledgements}
This work was supported by the University Grant Council (Contract No. AoE/P-04/08) of  the Government of HKSAR, National Natural Science Foundation of China with No.11374246 (J. Wang), NSERC of Canada (H. Guo). We thank CLUMEQ, CalcuQuebec and Compute Canada for computation facilities.

\newpage
\makeatother

\end{document}